\documentclass[journal]{IEEEtran}
\usepackage{amsmath}
\usepackage{amssymb}
\usepackage{cite}
\usepackage{graphicx}
\usepackage[dvipsnames,table,xcdraw]{xcolor}
\usepackage{colortbl}
\usepackage{caption}
\usepackage{subcaption}
\usepackage{amsthm}
\usepackage{soul}
\usepackage{hhline}
\usepackage[labelformat=simple]{subcaption}
\usepackage{multirow}
\usepackage[latin1]{inputenc}          

\begin{document}

\title{Power Control of Converters Connected via an L Filter to a Weak Grid. A Flatness-Based Approach}

\author{Sebastian Gomez Jorge$\dag$, Jorge~A.~Solsona$\dag$, Claudio A.  Busada$\dag$, Gerardo Tapia-Otaegui$\ddag$, Ana Susperregui$\ddag$ and M. Itsaso Martínez$\ddag$\\
\thanks{This work was supported in part by the Universidad Nacional del Sur (UNS), in part by Consejo Nacional de Investigaciones Cient\'ificas y T\'ecnicas (CONICET), in part by the Spanish Ministry of Science and Innovation (project code PID2020-115484RA-I00), FEDER Funds, EU, and in part by the Basque Government under research grant IT1644-22.
$\dag$ Authors are with the Instituto de Investigaciones en Ingenier\'ia El\'ectrica (IIIE), Universidad Nacional del Sur (UNS)-CONICET and Dpto. Ing. El\'ectrica y de Computadoras, UNS, Bah\'ia Blanca, Argentina.
$\ddag$ Authors are with the Department of Automatic Control and Systems Engineering, University of the Basque Country UPV/EHU, Faculty of Engineering--Gipuzkoa, Donostia 20018, Spain.
(e-mail: sebastian.gomezjorge@uns.edu.ar; jsolsona@uns.edu.ar; cbusada@uns.edu.ar; gerardo.tapia@ehu.eus; ana.susperregui@ehu.eus, mirenitsaso.martinez@ehu.eus).}
}

\maketitle

\begin{abstract}
In this article, a nonlinear strategy based on a flatness approach is used for controlling the instantaneous complex power supplied from the Point of Common Coupling (PCC) to a weak grid. To this end, the strategy introduced by the authors in \cite{10113842} considering a strong grid is robustified for avoiding system instability when the converter is connected to an unknown grid. The robustification method consists of including a notch filter that estimates the PCC voltage and using it to build the controller (i.e. the measured PCC voltage used to design the control strategy for a strong grid is replaced by the PCC voltage estimated with the notch filter). In addition, before designing the controller, the steady-state stability and safe operation limits when injecting complex instantaneous power to a grid of unknown impedance are analyzed. This analysis is independent of the control strategy, and applies to all power injection schemes.

Simulations are presented for showing the performance of the proposed controller in presence of a weak grid.
\end{abstract}


\section{Introduction}
Currently, most nations are considering decarbonizing energy generation in order to reduce the emission of greenhouse gases into the atmosphere. To meet this goal, the energy matrices of these countries are being modified to include electric power generators, whose primary source does not include gas or oil. In this context, the penetration of generators based on wind or photovoltaic energy stands out. These generators use power electronic converters to inject energy into the grid. There are different topologies for building grid-supporting or grid-forming converters  \cite{rathnayake2021grid,9552499,9866593,9652035}.
Among them, it is common to find a three-phase inverter with an inductive output filter coupled to the Point of Common Coupling (PCC) with the aim of injecting instantaneous complex power. Depending on the short circuit ratio, the line where the generator is connected could be weak or strong. When the grid is strong, it is possible to design the converter controller considering that the voltage value at the PCC is similar to the grid voltage value. However, a design of this type when the grid is weak can lead to instabilities in the system, so it is necessary to establish a robust design considering the value of the voltage at the PCC is different from that of the grid, and that the parameters of the line that interconnects them are no longer negligible.

It must be mentioned that the control strategy introduced in \cite{10113842} is nonlinear, and it allows to use a relatively small value of the DC-link capacitor. This controller presents better performance than those controllers designed by using linear techniques, since a coupled nonlinear control model is considered to design it. Note that, unlike this nonlinear proposal, linear controllers are designed by considering two decoupled loops, where a slow outer loop is used for controlling the DC-link voltage. Moreover, this kind of controllers are of limited performance, and only work well when the converter is affected by small disturbances and/or it must track slowly varying references \cite{10113842}. To overcome this drawback, many authors have proposed to use nonlinear controllers (see, among others, \cite{10113842} and several references therein). All controllers, linear and nonlinear, are designed based on the available information of the plant to be controlled, and the system's performance is strongly dependent on the good use of this information. The authors have shown that a grid-feeding converter can obtain excellent performance to large state variations and large disturbances by using nonlinear complex power control strategies  when the converter is connected to a strong grid (see \cite{10113842,TAPIAOTAEGUI2023451}).

However, it can be noted that sometimes the converter is connected to a grid with unknown parameters, where its resistance, inductance and voltage value differ from those assumed when designing the converter's controller. In this case, when power is delivered by the converter through the PCC, the whole system's performance can be widely deteriorated, even making its behavior unstable. Many papers can be found in the literature where researchers have considered some of these cases. These papers consider the converter connected to a weak grid with unknown inductance. Often, linear strategies are used for controlling the electrical variables, then only small-signal stability is guaranteed. A case where the influence on the stability of the unknown grid parameters was analyzed can be found in \cite{8242351}. Some robust controllers can be found in \cite{xu2017robust,8486047,8490668,wang2020robust}. Another control strategy was presented in \cite{9364526}. In \cite{9042561}, a controller based on model predictive control was also introduced. A power controller was proposed in \cite{khan2020single}, whereas a study of dynamic behavior in presence of a weak grid was presented in \cite{9530704}. The authors of \cite{9502732} show the behavior of a converter tied to a weak grid containing active damping.

It must be noted that grid-tied converters are used for injecting power at different levels, then they are connected to lines with several voltage values (i.e. High-Voltage, Medium-Voltage and Low-Voltage). Depending on the voltage level, these lines behave as resistive, inductive or a combination of both \cite{1600497}.

By considering the above-mentioned issues, the main goal in this work is to strengthen the controller that the authors have introduced in \cite{10113842} so that it can be used in the presence of a weak grid guaranteeing the system stability (i.e. connected to a grid with unknown parameters). It is considered that line resistance and line inductance are unknown, and that the grid voltage cannot be measured (although the PCC voltage is available).
Before designing the controller, the Steady-State (SS) stability limits of injecting complex power to the PCC, and the safe operation limits where the current is bounded, are analyzed. This analysis is independent of the control strategy, and applies to all complex power injection controllers.
Afterwards, the controller is built in two steps. First, a trajectory tracking instantaneous complex power strategy is designed, assuming a low impedance grid, as it is typically done for these controllers. Then, the transient stability of the proposal when in presence of grid impedance is analyzed. As it will be shown, for inductive grids, transient stability can be achieved by filtering the measured PCC voltage. To this end, the use of a notch filter is proposed, and a criterion to choose its gain in relation to the controller gains is developed.

\section{System's Model}
\begin{figure}[tb]
	\centering
	\includegraphics[width=1\linewidth]{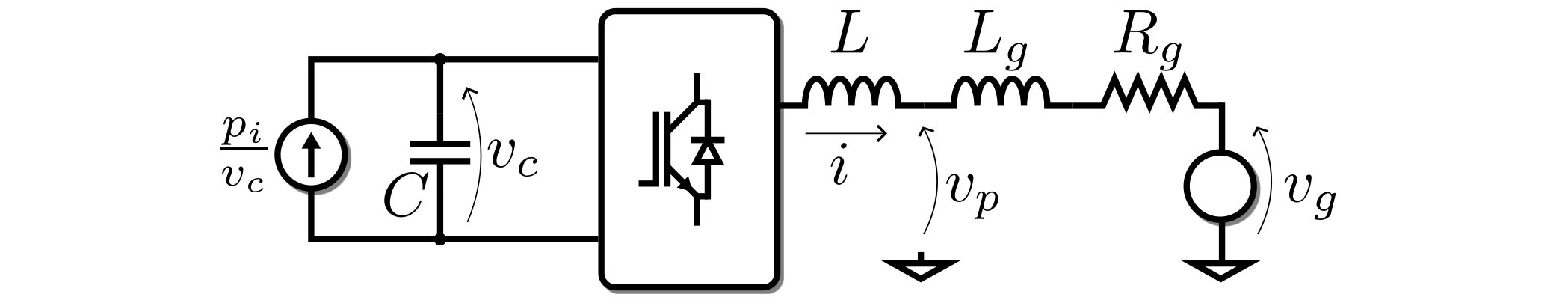}
	\caption{System's model using complex space vectors.}\label{fig:esquema}
\end{figure}
A three-phase grid-tied inverter with a DC-link capacitor fed from an input power source is depicted in Fig. \ref{fig:esquema}, where the three-phase variables are represented using complex space vectors. This system can be 
\begin{align}
	L\dot{i}&=v_c\mu-v_p,\label{eq:L di}\\
	C\dot{v}_c&=\frac{p_i}{v_c}-\Re\{\mu\tilde{i}\},\label{eq:C vC}\\
	(L+L_g)\dot{i}&=v_c\mu-v_g-R_gi,\label{eq:L+Lg di}
\end{align}
where $\tilde{.}$ denotes complex conjugate of the variable, $v_p$ is the PCC voltage, $i$ is the injected current, $v_c$ is the DC-link voltage, $\mu$ is the modulation index, $p_i$ is the input power source, $L$ is the inverter filter inductance, $L_g$ and $R_g$ are the grid inductance and resistance, respectively, and $v_g$ is the grid voltage. To simplify the study, $v_g$ is assumed to be a positive sequence signal of angular frequency $\omega$. As this is usually the dominant component of the grid voltage, the results obtained here also apply when harmonics and/or imbalance are present. Although the exact values of $L_g$ and $R_g$ are considered unknown, some degree of knowledge is required. Mainly, if the grid is inductive or resistive, as this drastically changes the active and reactive power limits for SS stability in weak grids, as it will be shown in Section \ref{sec:ss stability}.

Additionally, it is convenient to derive the algebraic expression of the PCC voltage, which will be used to study the transient stability in Section \ref{sec:transient stability midiendo vp}. From (\ref{eq:L di}) and (\ref{eq:L+Lg di}), this voltage results
\begin{align}
	v_p&=v_{p\alpha}+jv_{p\beta}=\frac{L_gv_c\mu+L(v_g+R_gi)}{L+L_g}=V_p e^{j\theta},\label{eq:vp}
\end{align}
where $V_p=|v_p|$ and $\theta=\arctan{\!\rm2}\{\Im\{v_p\},\Re\{v_p\}\}$.

\section{Steady-State Stability for Weak Grids}\label{sec:ss stability}
When the grid has impedance, there is a limit to the power that can be injected at the PCC while maintaining the stability of the system. To find this limit, the SS behavior of (\ref{eq:L di}) and (\ref{eq:L+Lg di}) is analyzed. By assuming a pure sinusoidal grid voltage with angular frequency $\omega$, which leads to $\dot{i}=j\omega i$ in SS, and using this result in (\ref{eq:L di}) and (\ref{eq:L+Lg di}), the following is obtained:
\begin{align}
	0&=(R_g+jX_g) |i|^2+v_g\tilde{i} -s,\label{eq:i ss}
\end{align}
where $X_g=\omega L_g$ and
\begin{align}
	s=v_p\tilde{i}=p+jq.\label{eq:s}
\end{align}
Solving this equation leads to
\begin{align}
	|i|^2&=\frac{2\,R_g\,p+2\,X_g\,q-|v_g|\sqrt{\lambda}+|v_g|^2 }{2\,(R_g^2 +X_g^2)},\label{eq:mag_i2}
\end{align}
where
\begin{align}
	 \lambda&=|v_g|^2\!\!-\!4X_g\!\left(\!\!\frac{X_gp^2}{|v_g|^2}\!-\!q\!\!\right)\!\!+\!4R_g\!\!\left(\!\frac{2X_gpq\!-\!R_gq^2}{|v_g|^2}\!+\!p\!\!\right)\label{eq:lambda}
\end{align}
must be greater or equal to zero for the solution to exist. This condition $\lambda\geq0$ is used to find the relationship between $p$ and $q$ for stable operation in SS.

In addition, the magnitude of the PCC voltage should be determined, since it is related to the control action limits of the inverter. From (\ref{eq:L di}) and (\ref{eq:L+Lg di}) evaluated in SS,
\begin{align}
	v_p=(R_g+jX_g)i+v_g.\label{eq:vp ss}
\end{align}
Solving for $i$ in (\ref{eq:i ss}) and using this result in (\ref{eq:vp ss}), yields
\begin{align}
	|v_p|^2&=R_gp+X_gq+\frac{|v_g|}{2}(|v_g|+\sqrt{\lambda}).\label{eq:|vp|^2 ss}
\end{align}

Some useful insights can be gained by considering the cases of purely inductive and purely resistive grids.

\subsection{Inductive Grids}
In the case of purely inductive grids ($R_g\rightarrow0$), assuming $\lambda\geq0$ in (\ref{eq:lambda}) yields the following relationship between $p$ and $q$ for stable operation in SS:
\begin{align}
	q&\geq\frac{X_g p^2}{|v_g|^2}-\frac{|v_g|^2}{4X_g}.\label{eq:q limite estable inductivo}
\end{align}
Moreover, keeping the magnitude of the current below or equal to its nominal value is also required for safe operation. Evaluating (\ref{eq:mag_i2}) and (\ref{eq:lambda}) for $R_g\rightarrow0$, and solving for $q$, the following limits for $q$ are found:
\begin{align}
	X_g|i|^2\!-\!\sqrt{|v_g|^2|i|^2\!-\!p^2}&\leq\!q\!\leq X_g|i|^2\!+\!\sqrt{|v_g|^2|i|^2\!-\!p^2},\label{eq:q safe zone inductivo}
\end{align}
where
\begin{align}
	|p|&\leq|v_g||i|\label{eq:lim p}
\end{align}
must also hold. Notice that
the limits of (\ref{eq:q safe zone inductivo}) contain the stable operation zone of (\ref{eq:q limite estable inductivo}), and that (\ref{eq:q limite estable inductivo}) only touches the bounds of (\ref{eq:q safe zone inductivo}) at one point, which is found by equating the right side of (\ref{eq:q limite estable inductivo}) with the term to the left of the first inequality of (\ref{eq:q safe zone inductivo}), resulting
\begin{align}
	X_g&=\frac{|v_g|^2}{2\sqrt{|v_g|^2|i|^2\!-\!p^2}}.
\end{align}

The next limit that should be found is the maximum control action. If Space Vector Modulation (SVM) is used, then $|\mu|\leq1/\sqrt{2}$ (see  \cite{TAPIAOTAEGUI2023451}). Evaluating (\ref{eq:|vp|^2 ss}) for $R_g\rightarrow0$, equating the result to the square of the maximum control action $(v_c|\mu_{max}|)^2=v_c^2/2$, and then solving for $q$, yields
\begin{align}
	q\leq\frac{v_c^2-\sqrt{2}\sqrt{v_c^2|v_g|^2-2X_g^2p^2}}{2X_g},\label{eq:limite control inductivo}
\end{align}
which gives a rough estimate of the $q$ that saturates the control action. This limit is approximate, because this approach neglects the voltage drop of the inverter filter.

\begin{figure}[tb]
	\centering
	\includegraphics[width=1\linewidth]{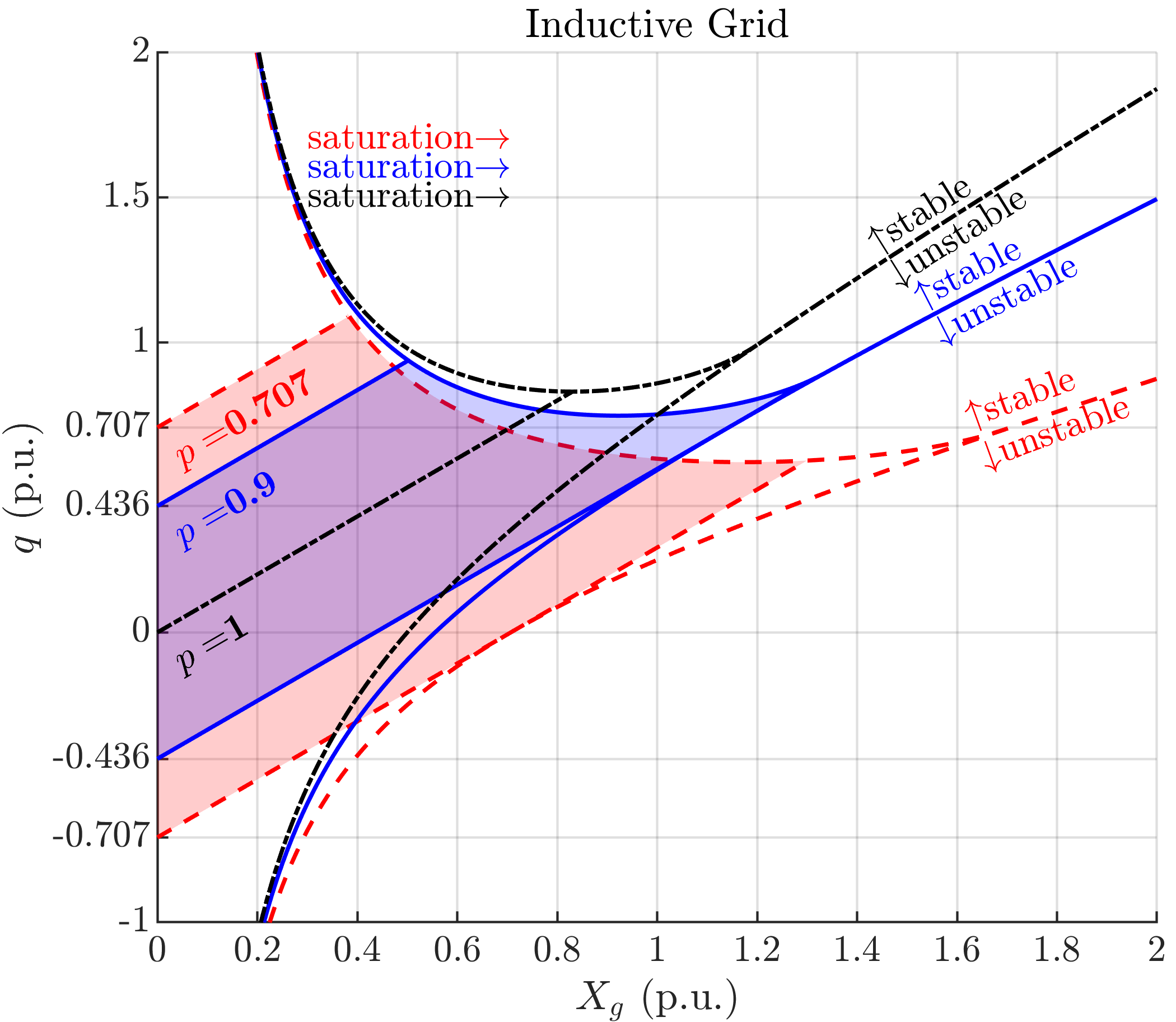}
	\caption{Reactive power limits for $|i|\leq1$, zone of stable operation and control action limit in SS for inductive grids.}\label{fig:limite q inductive grid}
\end{figure}
Figure \ref{fig:limite q inductive grid} depicts a plot of (\ref{eq:q limite estable inductivo}), (\ref{eq:q safe zone inductivo}) and (\ref{eq:limite control inductivo}) for three values of $p$. In this case, (\ref{eq:limite control inductivo}) is plotted for $v_c=1.3\sqrt{2}$ ($1.3$ times the voltage required to generate nominal grid voltage using SVM). For $p=0.707$, the magnitude of the current is $|i|\leq1$ for $q$ within the area between the dashed lines, while stable operation is attained for $q$ above the dashed line curve of the bottom. Additionally, the control action saturates above the dashed curve at the top. As $p$ is increased, the area where $|i|\leq1$ becomes smaller, as it can be seen for $p=0.9$, where $|i|\leq1$ for $q$ inside the area between the solid lines, and stable operation is obtained above the solid line curve. The limit case is $p=1$, where $|i|=1$ on the dash-dot line, and it is larger elsewhere (stable operation for $q$ above the dash-dot curve).

\subsection{Resistive Grids}
\begin{figure}[tb]
	\centering
	\includegraphics[width=1\linewidth]{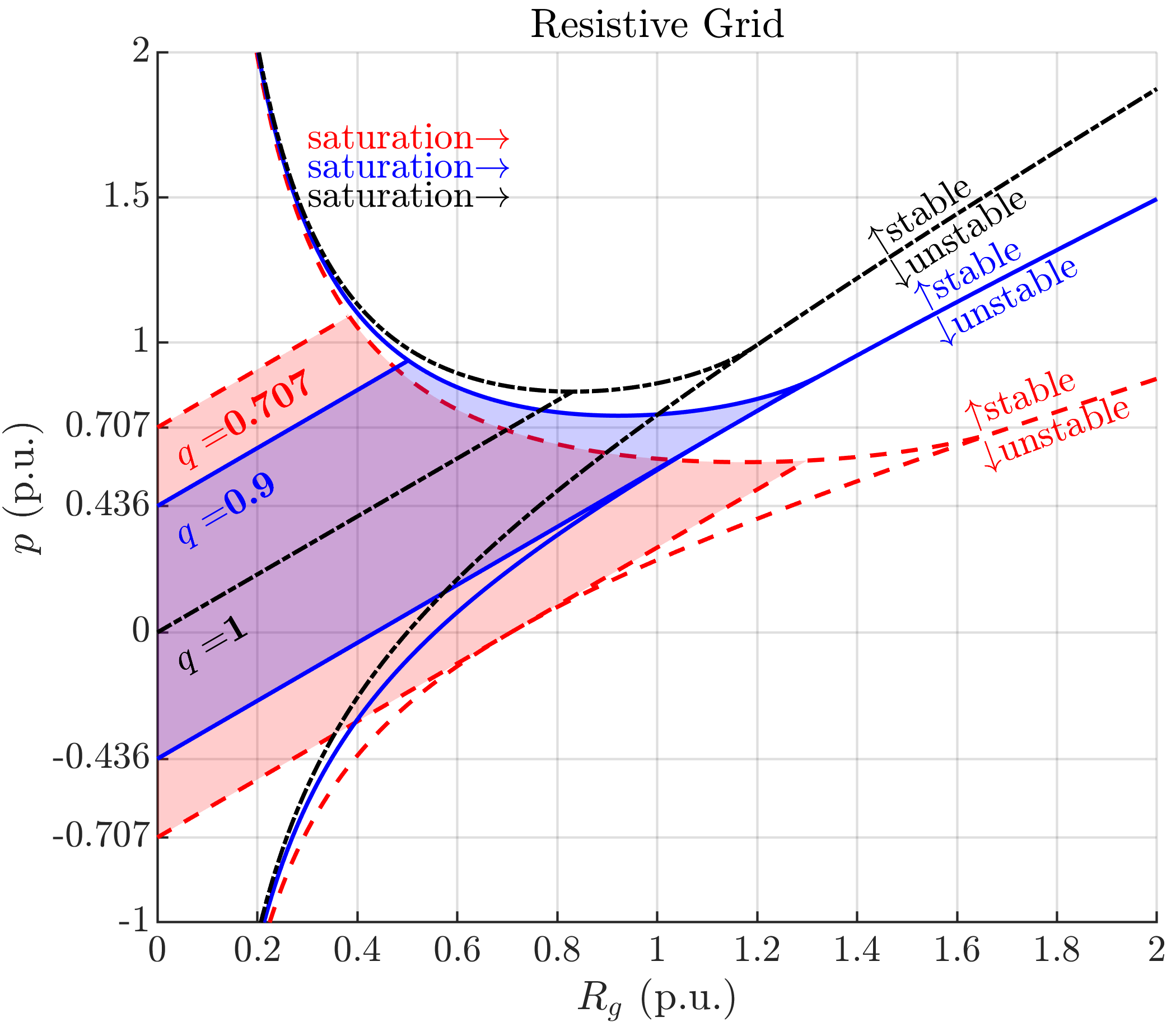}
	\caption{Active power limits for $|i|\leq1$, zone of stable operation
		and control action limit in SS for resistive grids.}\label{fig:limite p resistive grid}
\end{figure}
In the case of purely resistive grids ($X_g\rightarrow0$), from (\ref{eq:lambda}) the stable operation region is given by
\begin{align}
	p&\geq\frac{R_gq^2}{|v_g|^2}-\frac{|v_g|^2}{4R_g},\label{eq:p limite estable resistivo}
\end{align}
and the safe area of operation where $|i|\leq1$, obtained from (\ref{eq:mag_i2})-(\ref{eq:lambda}), results
\begin{align}
	R_g|i|^2\!-\!\sqrt{|v_g|^2|i|^2\!-\!q^2}&\leq\!p\!\leq R_g|i|^2\!+\!\sqrt{|v_g|^2|i|^2\!-\!q^2}.\label{eq:p safe zone resistivo}
\end{align}
Additionally, the control action saturation limit obtained from (\ref{eq:|vp|^2 ss}) results
\begin{align}
	p\leq\frac{v_c^2-\sqrt{2}\sqrt{v_c^2|v_g|^2-2R_g^2q^2}}{2R_g}.\label{eq:limite control resistivo}
\end{align}
Comparing (\ref{eq:p limite estable resistivo})--(\ref{eq:limite control resistivo}) with (\ref{eq:q limite estable inductivo})-(\ref{eq:q safe zone inductivo}) and (\ref{eq:limite control inductivo}), it is clear that the resistive grid case is analogous to the inductive grid case. As a result, the plot of (\ref{eq:p limite estable resistivo})--(\ref{eq:limite control resistivo}), shown in Fig. \ref{fig:limite p resistive grid}, is identical to the plot of Fig. \ref{fig:limite q inductive grid} (renaming the axes and swapping $p$ for $q$). This plot shows the safe area of operation for three values of $q$, along with the stability and control action limits in each case.

\section{Control Based on a Flatness Approach}
The system (\ref{eq:L di})-(\ref{eq:C vC}) can be feedback linearized using the change of variables proposed in \cite{10113842}. To ease the analysis, a simplified version of that controller is summarized here, where no dead time disturbance or grid voltage harmonics are considered. Assuming $L_g=R_g=0$ and $v_p=v_g$ known, the Full Feedback Linearization (FFL) of system (\ref{eq:L di})-(\ref{eq:C vC}) can be obtained in terms of the complex energy $\xi_1$ and the complex power balance $\xi_2$, which are defined below. Let
\begin{align}
	\xi_1&=\frac{1}{2}(L|i|^2+Cv_c^2)+j\eta,\label{eq:xi1}
\end{align}
be the complex energy of the system, where $\dot{\eta}=\Im\{v_p\tilde{i}\}=q$ is the reactive power at the PCC. Differentiating (\ref{eq:xi1}) with respect to time, and replacing with (\ref{eq:L di})-(\ref{eq:C vC}), yields the complex power balance $\xi_2$, that results
\begin{align}
	\dot{\xi}_1&=p_i-\tilde{v}_pi=p_i-p+jq=\xi_2,\label{eq:xi2}
\end{align}
where $p=\Re\{v_p\tilde{i}\}$ is the active power at the PCC.
Differentiating (\ref{eq:xi2}) with respect to time, the auxiliary control action $u$ of the linearized system is obtained
\begin{align}
	\dot{\xi}_2&=\dot{p}_i-\tilde{v}_p\dot{i}-\dot{\tilde{v}}_pi=u.\label{eq:dxi2}
\end{align}
A controller for computing $u$ in order to control the resulting linear system (\ref{eq:xi2})-(\ref{eq:dxi2}) can now be designed using linear control techniques. Then from (\ref{eq:L di}) and (\ref{eq:dxi2}), the modulation index that must actually be applied to the original system is computed as
\begin{align}
	\mu&=\frac{L(\dot{p}_i-u-\dot{\tilde{v}}_pi)+|v_p|^2}{\tilde{v}_pv_c}.\label{eq:mu}
\end{align}
The computation of (\ref{eq:mu}) requires knowledge of $\dot{v}_p$. Without loss of generality, for the design of the controller, it will be assumed that the grid is strong ($v_p\simeq v_g$), and therefore that $\dot{v}_p=j\omega v_p$, so (\ref{eq:mu}) is computed as
\begin{align}
	\mu&=\frac{L(\dot{p}_i-u+j\omega\tilde{v}_pi)+|v_p|^2}{\tilde{v}_pv_c}.\label{eq:mu implementada}
\end{align}
If more information about the grid is available, such as harmonic content and imbalance, this can be used to improve the design of the controller. Additionally, note that the computation of $\dot{p}_i$ in (\ref{eq:mu implementada}) can be avoided by using a trajectory tracking controller, as it will be done here. Also, to avoid measuring $p_i$, the observer described in \cite{10113842} can be used.

\subsection{Controller Design}
A full state feedback controller along with trajectory tracking is used
\begin{align}
	u&=\dot{\xi}_{2r}-k_1e_{\xi_1}-k_2e_{\xi_2}-k_3y,\label{eq:u}\\
	\dot{y}&=e_{\xi_1},\label{eq:dy}
\end{align}
where $e_{\xi_1}=\xi_1-\xi_{1r}$, $e_{\xi_2}=\xi_2-\xi_{2r}$, with $\xi_{1r}$, $\xi_{2r}$ being the references of $\xi_1$ and $\xi_2$, respectively, $[k_1\ k_2\ k_3]\in\mathbb{R}^+$ are constants that define the feedback dynamics, and $y$ is an integral state. These constants can be computed from the error dynamics obtained from (\ref{eq:xi2})-(\ref{eq:dxi2}) and (\ref{eq:u})-(\ref{eq:dy}), ${\bf\dot{e}_\xi}={\bf A}{\bf e_\xi}+{\bf B}(u-\dot{\xi}_{2r})$, where
\begin{align}
	{\bf A}&=
	\begin{bmatrix}
		0&1&0\\
		0&0&0\\
		1&0&0
	\end{bmatrix};\
	{\bf B}=\begin{bmatrix}
		0\\
		1\\
		0
	\end{bmatrix},
\end{align}
and ${\bf e_\xi}=[e_{\xi_1}\ e_{\xi_2}\ y]^T$, using ${\bf A}$, ${\bf B}$ and any pole placement technique, such as Ackerman's method.

\subsection{References}
Let $v_{cr}$ and $q_r$, and their time derivatives, be arbitrary signals that represent the references for $v_c$ and $q$, respectively. Then, from (\ref{eq:xi1})-(\ref{eq:xi2})
\begin{align}
	\xi_{1r}&=\frac{1}{2}(L|i_r|^2+Cv_{cr}^2)+j\eta_r,\\
	\xi_{2r}&=p_i-p_r+jq_r,\\
	\dot{\xi}_{2r}&=\dot{p}_i-\dot{p}_r+j\dot{q}_r,
\end{align}
where $\dot{\eta}_r=q_r$, $|i_r|^2=(p_r^2+q_r^2)/|v_p|^2,$ and
\begin{align}
	\dot{p}_r&=\frac{|v_p|^2(p_i-p_r-C\dot{v}_{cr}v_{cr})-L\dot{q}_rq_r}{L(|p_r|+\delta_p)},\label{eq:dpr}
\end{align}
with $\delta_p>0$ an arbitrary constant to avoid zero division. The derivation of (\ref{eq:dpr}) is detailed in \cite{10113842}, and assumes that $|v_p|$ is slowly varying.

\subsection{Transient Stability Measuring $v_p$}\label{sec:transient stability midiendo vp}
The stability of the whole system (\ref{eq:L di})--(\ref{eq:L+Lg di}) (i.e. considering $R_g$, $L_g\neq0$), will be analyzed.
From (\ref{eq:L+Lg di}), (\ref{eq:xi2})-(\ref{eq:dxi2}) and (\ref{eq:mu implementada})--(\ref{eq:dy}), the error dynamics of the feedback linearized system are given by
\begin{align}
	\begin{bmatrix}
		\dot{e}_{\xi_1}\\
		\dot{e}_{\xi_2}\\
		\dot{y}
	\end{bmatrix}&=
	\underbrace{\begin{bmatrix}
		0&1&0\\
		-K_1&-K_2&-K_3\\
		1&0&0
	\end{bmatrix}}_{\bf A_{cl}}
	\begin{bmatrix}
		e_{\xi_1}\\
		e_{\xi_2}\\
		y
	\end{bmatrix}+
	\begin{bmatrix}
		0\\
		1\\
		0
	\end{bmatrix}\nu(v_p),
\end{align}
where
\begin{align}
	K_1\!\!&=\!\!\frac{Lk_1}{L\!+\!L_g};\,
	K_2\!\!=\!\!\frac{R_g\!+\!L(k_2\!-\!j\omega )}{L\!+\!L_g}\!-\!\frac{\dot{\tilde{v}}_p}{\tilde{v}_p};\,
	K_3\!\!=\!\!\frac{Lk_3}{L\!+\!L_g},
\end{align}
and
\begin{align}
	\nu(v_p)&=\frac{L_g(\dot{p}_i-\dot{\xi}_{2r})+\tilde{v}_p(v_g-v_p)}{L+L_g}\nonumber\\
	&+\frac{\left[R_g-j\omega L-\frac{\dot{\tilde{v}}_p}{\tilde{v}_p}(L+L_g)\right](p_i-\xi_{2r})}{L+L_g}
\end{align}
can be considered as an input that does not affect the stability. The stability of the closed-loop system can be analyzed from the characteristic polynomial of $\bf A_{cl}$,	$f(\lambda)=\lambda^3+K_2\lambda^2+K_1\lambda+K_3$ by
using the generalized Routh-Hurwitz criterion \cite{HASTIR2023100235}, from which
\begin{align}
	&K_{2r}>0,\label{eq:Routh-Hurwitz condicion 1}\\
	&-K_{2r}(K_3-K_1K_{2r})>0,\\
	&K_3K_{2r}^2(K_3-K_1K_{2r})^2-K_3^2K_{2i}^2K_{2r}^3>0,\label{eq:Routh-Hurwitz condicion 3}
\end{align}
must hold for the system to be stable, where
\begin{align}
	K_{2r}&=\frac{R_g+Lk_2}{L+L_g}-\frac{\dot{V}_p}{V_p};\ K_{2i}=-\frac{\omega L}{L+L_g}+\dot{\theta},\label{eq:K2r}
\end{align}
with $V_p$ and $\theta$ defined in (\ref{eq:vp}). Notice the issue highlighted by (\ref{eq:K2r}), which makes it very hard for (\ref{eq:Routh-Hurwitz condicion 1}) to hold. On the one hand, $k_2$ should be ``large'' to keep $K_{2r}>0$, even when in presence of sudden increases of $V_p$. But from (\ref{eq:vp}), $\dot{V}_p\propto\dot{\mu}$, which means the larger the controller gains, the larger $\dot{V}_p$. Even ``soft'' reference changes, such as slowly increasing the DC-link voltage or the reactive power, might lead to $\dot{V}_p$ being large enough to destabilize the system.

In order to avoid this issue, it is proposed to filter $v_p$, effectively eliminating the coupling of (\ref{eq:K2r}) with $\dot{\mu}$, leading to a much easier strategy to stabilize the system. Notice that, although the issue would not be present in ``purely'' resistive grids, since the coupling with $\dot{\mu}$ would not be present in (\ref{eq:vp}), it is still recommended to apply the proposed filtering scheme, as in reality there will always be some inductive component in the grid.

\subsection{Transient Stability Filtering $v_p$}
Let $\hat{v}_p=\hat{v}_{p\alpha}+j\hat{v}_{p\beta}=\hat{V}_pe^{j\hat{\theta}}$ be a filtered version of $v_p$, and from (\ref{eq:xi1})-(\ref{eq:xi2}) and (\ref{eq:u})-(\ref{eq:dy}), define
\begin{align}
	\hat{\xi}_1&=\frac{1}{2}(L|i|^2+Cv_c^2)+j\hat{\eta},\label{eq:xi1_est}\\
	\hat{\xi}_2&=p_i-\tilde{\hat{v}}_pi=p_i-\hat{p}+j\hat{q},\label{eq:xi2_est}\\
	\hat{u}&=\dot{\xi}_{2r}-k_1\hat{e}_{\xi_1}-k_2\hat{e}_{\xi_2}-k_3\hat{y},\label{eq:u_est}\\
	\dot{\hat{y}}&=\hat{e}_{\xi_1},\label{eq:dy_est}
\end{align}
where $\dot{\hat{\eta}}=\Im\{\hat{v}_p \tilde{i}\}=\hat{q}$, $\hat{p}=\Re\{\hat{v}_p\tilde{i}\}$, $\hat{e}_{\xi_1}=\hat{\xi}_1-\xi_{1r}$ and $\hat{e}_{\xi_2}=\hat{\xi}_2-\xi_{2r}$. Now, differentiating (\ref{eq:xi1_est})-(\ref{eq:xi2_est}) with respect to time yields
\begin{align}
	\dot{\hat{\xi}}_1&=\hat{\xi}_2-\Re\{\tilde{i}e_{v_p}\},\\
	\dot{\hat{\xi}}_2&=\underbrace{\dot{p}_i\!+\!j\omega \tilde{\hat{v}}_pi\!-\!\tilde{\hat{v}}_p\left(\dot{i}\!+\!\frac{e_{v_p}}{L}\right)}_{\hat{u}}\!-\!(j\omega \tilde{\hat{v}}_p\!+\!\dot{\tilde{\hat{v}}}_p)i\!+\!\tilde{\hat{v}}_p\frac{e_{v_p}}{L}.\label{eq:dxi2_est}
\end{align}
where $e_{v_p}=v_p-\hat{v}_p$. Then, from (\ref{eq:L di}) and (\ref{eq:dxi2_est}), the modulation index results
\begin{align}
	\mu&=\frac{L(\dot{p}_i+j\omega \tilde{\hat{v}}_pi-\hat{u})+|\hat{v}_p|^2}{v_c\tilde{\hat{v}}_p}.\label{eq:mu_est}
\end{align}
Notice that the variables have been carefully selected in order to avoid requiring $v_p$ in their computation. The control is implemented by computing (\ref{eq:xi1_est})--(\ref{eq:dy_est}), and then (\ref{eq:mu_est}).

From (\ref{eq:L+Lg di}), (\ref{eq:xi1_est})--(\ref{eq:dy_est}) and (\ref{eq:mu_est}), the error dynamics of the feedback linearized system is given by
\begin{align}
	\begin{bmatrix}
		\dot{\hat{e}}_{\xi_1}\\
		\dot{\hat{e}}_{\xi_2}\\
		\dot{\hat{y}}
	\end{bmatrix}\!\!\!&=\!\!\!
	\underbrace{\begin{bmatrix}
			0&1&0\\
			-K_1&-\hat{K}_2&-K_3\\
			1&0&0
	\end{bmatrix}}_{\bf \hat{A}_{cl}}
	\!\!\!\begin{bmatrix}
		\hat{e}_{\xi_1}\\
		\hat{e}_{\xi_2}\\
		\hat{y}
	\end{bmatrix}\!\!\!+\!\!\!
	\begin{bmatrix}
		0&-1\\
		1&0\\
		0&0
	\end{bmatrix}\!\!\!
	\begin{bmatrix}
		\nu(\hat{v}_p)\\
		\Re\{\tilde{i}e_{v_p}\}
	\end{bmatrix}\!\!,
\end{align}
where
\begin{align}
	\hat{K}_2&=\frac{R_g+L(k_2-j\omega )}{L+L_g}-\frac{\dot{\tilde{\hat{v}}}_p}{\tilde{\hat{v}}_p}=\hat{K}_{2r}+j\hat{K}_{2i}.\label{eq:K2_est}
\end{align}
At this point, it is necessary to define the filter dynamics in order to find a stability criterion. Here, it is proposed to filter $v_p$ using a notch filter tuned to the grid angular frequency $\omega$,
\begin{align}
     \dot{\hat{v}}_p&=j\omega\hat{v}_p+\kappa (v_p-\hat{v}_p),\label{eq:dvp_est}
\end{align}
where $\kappa=\kappa_r+j\kappa_i$, with $[\kappa_r\ \kappa_i]\in\mathbb{R}^+$, is a constant that determines the filter settling time and damping. Replacing (\ref{eq:dvp_est}) in 
\begin{align}
	 \hat{K}_{2r}&=\frac{R_g\!+\!Lk_2}{L\!+\!L_g}\!-\!\kappa_r\left[\frac{V_p}{\hat{V}_p}\cos(e_\theta)\!-\!1\right]\!\!+\!\kappa_i\frac{V_p}{\hat{V}_p}\sin(e_\theta),\label{eq:K2r_est}\\
	\hat{K}_{2i}&=\frac{\omega L_g}{L\!+\!L_g}\!+\!\kappa_i\left[\frac{V_p}{\hat{V}_p}\cos(e_\theta)\!-\!1\right]\!\!+\!\kappa_r\frac{V_p}{\hat{V}_p}\sin(e_\theta),\label{eq:K2i_est}
\end{align}
where $e_\theta=\theta-\hat{\theta}$. Then, by defining bounds for $e_\theta$ and $V_p/\hat{V}_p$, these expressions can be used to evaluate (\ref{eq:Routh-Hurwitz condicion 1})--(\ref{eq:Routh-Hurwitz condicion 3}) and determine the stability for given conditions. However, some insight can be gained by doing some simplifications.

The expressions (\ref{eq:K2r_est})-(\ref{eq:K2i_est}) can be simplified by designing the filter with $\kappa_i=0$, and assuming $-0.5\pi\leq e_\theta\leq0.5\pi$. Using these assumptions in (\ref{eq:K2r_est})-(\ref{eq:K2i_est}) along with (\ref{eq:Routh-Hurwitz condicion 1})--(\ref{eq:Routh-Hurwitz condicion 3}), the following conservative conditions must hold for stability:
\begin{align}
	\hat{K}_{2r}-\frac{k_3}{k_1}\geq\frac{R_g+Lk_2}{L+L_g}-\kappa_r\frac{V_p-\hat{V}_p}{\hat{V}_p}-\frac{k_3}{k_1}&>0,\label{eq:condicion conservativa}\\
	\left(\hat{K}_{2r}-\frac{k_3}{k_1}\right)^2-\frac{K_3}{K_1^2}\left(\frac{\omega L_g}{L+L_g}+\kappa_r\frac{V_p}{\hat{V}_p}\right)^2\hat{K}_{2r}&>0.
\end{align}
Although there may seem not particularly useful, the important result is in (\ref{eq:condicion conservativa}), where it can be deduced that $k_2>>\kappa_r$ is a sufficient condition for stability. In other worlds, the rule of the thumb is that the convergence speed of the controller must be faster than that of the filter, which implies that, in order to gain robustness, it is necessary to sacrifice dynamic performance.

\section{Simulation Results}
To test the proposed stabilization method for inductive grids, the system is simulated under high grid impedance values. The per unit parameters of the system are as follows: the DC-link voltage reference is set to $v_{cr}=1.3\sqrt{2}$ ($1.3$ times the voltage required to generate nominal grid voltage using SVM), $L=0.02/\omega$, with $\omega=2\pi 50\,$rad/s, and $C=48\mu Z_b$, with $Z_b$ the base impedance. The gains $k_1$--$k_3$ are computed to obtain three real poles with settling times of $1$, $1.1$ and $20\,$ms, respectively. This yields $k_1=21.25e6$, $k_2=9011$ and $k_3=4424e6$. The gain of the filter is computed to obtain a complex pole with settling time of $50\,$ms, resulting $\kappa=92$.

\begin{figure}[tb]
	\centering
	\includegraphics[width=1\linewidth]{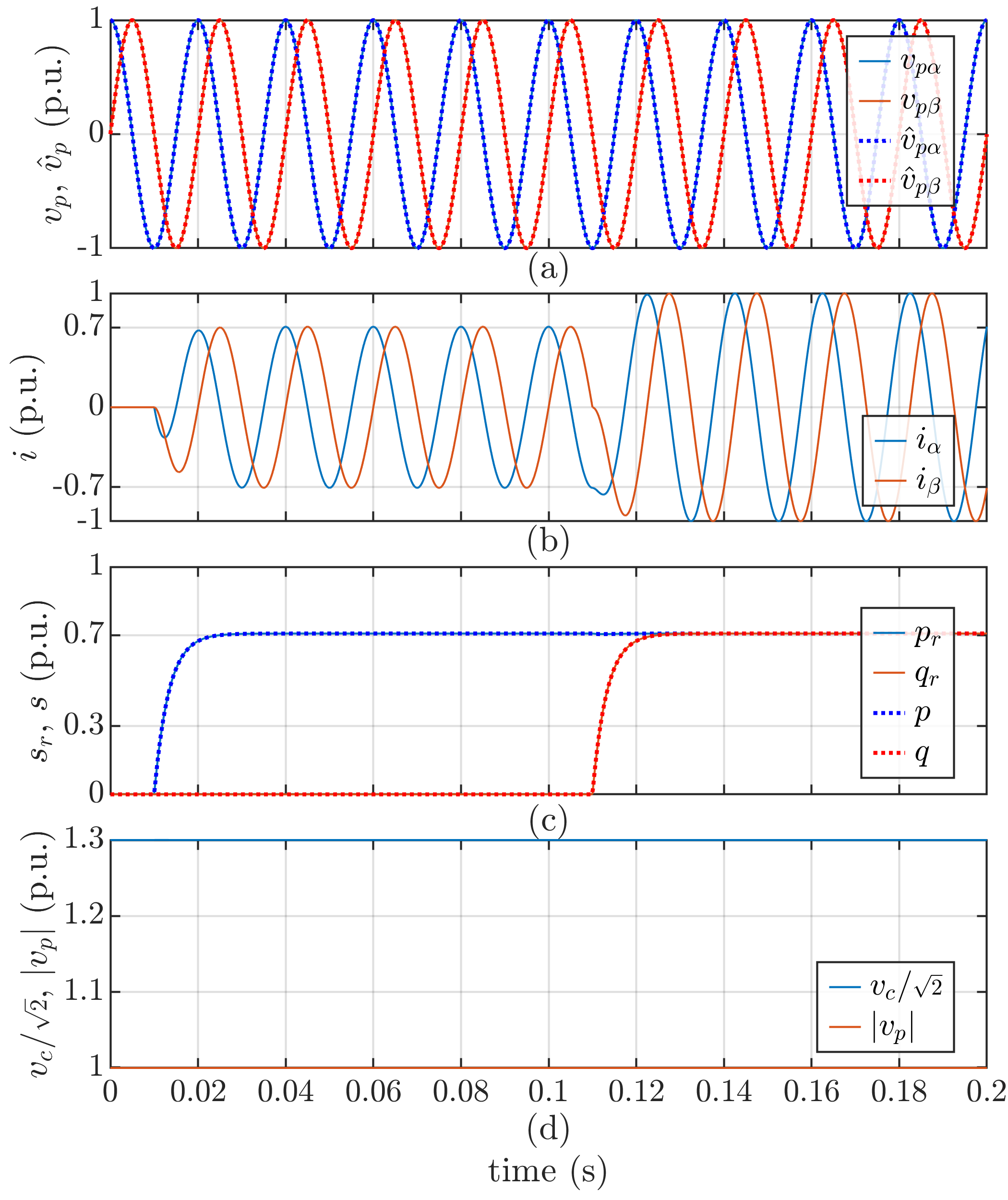}
	\caption{Strong grid simulation. (a) PCC voltage and filtered PCC voltage. (b) Injected current. (c) Reference power and injected power. (d) Scaled DC-link voltage and magnitude of PCC voltage.}\label{fig:sim red fuerte}
\end{figure}
For reference, Fig. \ref{fig:sim red fuerte} shows simulation results of the proposed control scheme with filtered $v_p$ against a strong grid ($R_g=L_g=0$). Figure \ref{fig:sim red fuerte}(a) shows the PCC voltage along with the filtered PCC voltage, Fig. \ref{fig:sim red fuerte}(b) shows the current, and Fig. \ref{fig:sim red fuerte}(c) shows the reference complex power along with the injected complex power. Finaly, Fig. \ref{fig:sim red fuerte}(d) shows the DC-link voltage, normalized to $1/\sqrt{2}$ along with the magnitude of the PCC voltage. As long as $|v_p|<v_c/\sqrt{2}$, the control action will not saturate.

In this simulation, first the input power is increased to $p_i=0.707$ at $t=0.01\,$s, and then the reactive power reference is increased to $q_r=0.707$ at $t=0.11\,$s. As it can be seen, the power references are perfectly tracked in this scenario.

\begin{figure}[tb]
	\centering
	\includegraphics[width=1\linewidth]{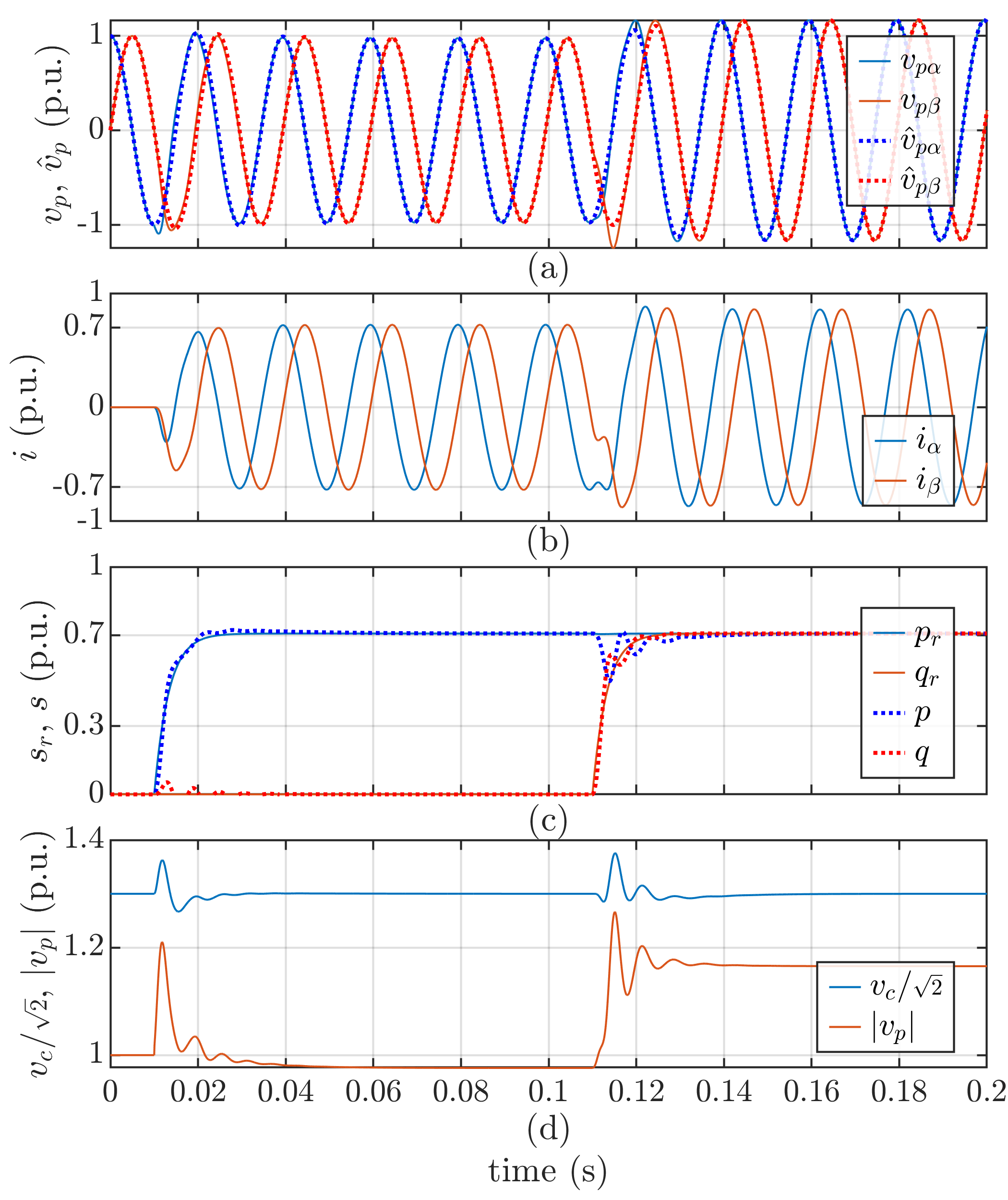}
	\caption{Weak inductive grid simulation. (a) PCC voltage and filtered PCC voltage. (b) Injected current. (c) Reference power and injected power. (d) Scaled DC-link voltage and magnitude of PCC voltage.}\label{fig:sim red Xg 0.3}
\end{figure}
To test the performance to weak inductive grids, a simulation with $X_g=0.3$ and $R_g=0$ is performed in Fig. \ref{fig:sim red Xg 0.3}, using the same power reference profiles as in Fig. \ref{fig:sim red fuerte}. In this case, there is coupling between $p$ and $q$ during the transients, which is to be expected due to the effect of the injected power on the PCC voltage. This coupling extinguishes with the notch filter dynamics. Figure \ref{fig:sim red Xg 0.3}(a) also shows that the PCC voltage magnitude is most affected by the reactive power, which is a known result for inductive grids. The magnitude of this voltage could be controlled by a secondary loop, but this controller escapes the scope of this paper. Notice also that, although not shown in the simulation, negative reactive power would decrease the magnitude of the PCC voltage, and from Fig. \ref{fig:limite q inductive grid}, the minimum stable reactive power for $X_g=0.3$ is $q\simeq-0.4$. This asymmetry in the reactive power limits implies that care must be taken when designing a secondary PCC voltage control-loop, as it is easy to wander into the unstable region of operation. Figure \ref{fig:sim red Xg 0.3}(d) shows that although $|v_p|$ increases, it does not go above the maximum control action, and therefore there is no control action saturation.

\section{Conclusions}
 A method for robustifying a nonlinear controller used to control the instantaneous complex power injected to an unknown grid was introduced. The power is injected via a grid-feeding inverter connected to a grid with unknown parameters through an L filter. It was demonstrated that fast variations in power reference cannot be tracked by a controller designed by considering a strong grid (i.e. by assuming the PCC and grid voltages are equal and the grid impedance is neglected), because the abrupt PCC voltage change necessary to track the power brings the system to instability when the converter is connected to a weak grid. In order to re-stabilize the system, it is needed to sacrifice transient performance by diminishing the rate of change of the PCC voltage to be fed back. This can be done in different ways. Here, the inclusion of a notch filter was proposed to obtain an estimated PCC voltage with a reduced value of rate of change, and this estimated value replaced the measured PCC voltage in the nonlinear control law. Moreover, it was shown how this filter should be calculated for guaranteeing the stability of the whole system. In addition, the limits of the injected power at the PCC to keep the whole system stable were calculated.

\bibliographystyle{IEEEtran}
\bibliography{biblio}

\end{document}